\documentclass[%
 reprint,
showkeys,
 amsmath,amssymb,
 aps,
]{revtex4-1}

\usepackage{graphicx}
\usepackage{dcolumn}
\usepackage{bm}

\usepackage{lipsum}
\usepackage{multirow}
\usepackage{graphicx}
\usepackage{dcolumn}
\usepackage{bm}

\usepackage[utf8]{inputenc}
\usepackage[T1]{fontenc}
\usepackage{mathptmx}

\usepackage{mathtools}
\usepackage{ mathrsfs }

\usepackage[dvipsnames]{xcolor}
\usepackage{color}
\usepackage{hyperref}
\hypersetup{colorlinks=true,
            linkcolor = NavyBlue,
            citecolor = NavyBlue,
            urlcolor = NavyBlue}
\usepackage[capitalize]{cleveref}

\graphicspath{{./Figures/}}

\newcommand{\e}{\epsilon}

\newcommand{\boldA}{\mathbf{A}}

\newcommand{\boldD}{\mathbf{D}}

\newcommand{\N}{\mathbb N}

\newcommand{\R}{\mathbb R}

\newcommand{\x}{\mathbf{x}}

\newcommand{\y}{\mathbf{y}}
\newcommand{\Z}{\mathbb Z}

\newcommand{\pers}{\textrm{pers}}
\newcommand{\maxpers}{\textrm{maxpers}}
\newcommand{\len}{\mathrm{len}}

\begin{document}


\title{Persistent Homology of Complex Networks for Dynamic State Detection}

\author{Audun Myers}
 \affiliation{Department of Mechanical Engineering, \\ Michigan State University.}

\author{Elizabeth Munch}%
\affiliation{Dept.~of Computational Mathematics, Science, and Engineering; \\ Dept.~of Mathematics; \\ Michigan State University.}

\author{Firas A.~Khasawneh}
 \affiliation{Department of Mechanical Engineering, \\ Michigan State University.\\}

\date{\today}

\begin{abstract}
In this paper we develop a novel Topological Data Analysis (TDA) approach for studying graph representations of time series of dynamical systems.
Specifically, we show how persistent homology, a tool from TDA, can be used to yield a compressed, multi-scale representation of the graph that can distinguish between dynamic states such as periodic and chaotic behavior.
We show the approach for two graph constructions obtained from the time series.
In the first approach the time series is embedded into a point cloud which is then used to construct an undirected $k$-nearest neighbor graph.
The second construct relies on the recently developed ordinal partition framework.
In either case, a pairwise distance matrix is then calculated using the shortest path between the graph’s nodes, and this matrix is utilized to define a filtration of a simplicial complex that enables tracking the changes in homology classes over the course of the filtration.
These changes are summarized in a persistence diagram—a two-dimensional summary of changes in the topological features.
We then extract existing as well as new geometric and entropy point summaries from the persistence diagram and compare to other commonly used network characteristics.
Our results show that persistence-based point summaries yield a clearer distinction of the dynamic behavior and are more robust to noise than existing graph-based scores, especially when  combined with ordinal graphs.
\end{abstract}

\maketitle


\section{Introduction}
\label{sec:intro}
\begin{figure}
    \centering
    \includegraphics[scale = 4]{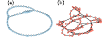}
    \caption{Comparison between ordinal partition networks generated from $x$-solution of R$\ddot{\rm o}$ssler system for both periodic (a) and chaotic (b) time series. }
    \label{fig:networks_PeriodicVsChaotic}
\end{figure}

\begin{figure*}
    \centering
    \includegraphics[width = \textwidth]{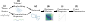}
    \caption{Outline of method: a time series (a) is embedded (b) into either $\mathbb{R}^n$ space using Takens' embedding or segmenting into a set of permutations. From these two representations, an undirected, unweighted network (c) is formed by either applying a $k^{th}$ nearest neighbors algorithm or by setting each permutation state as a node.
    The distance matrix (d) is calculated using the shortest path between all nodes.
    The persistence diagram (e) is generated by applying persistent homology to the distance matrix.
    Finally, one of several point summaries (f) are used to extract information from the persistence diagram. }
    \label{fig:MethodOutline}
\end{figure*}

There has been extensive work on understanding the behavior of the underlying dynamical system given only a time series \cite{Kantz2004,Bradley2015}.
The revolutionary work of Takens \cite{Takens1981} extended by Sauer et al.~\cite{Sauer1991} showed that, given most choices of parameters, the state-space of the dynamical system can be reconstructed through the Takens'embedding.
Computationally, this arises as the following procedure.
Given a time series $[x_1,\cdots,x_n]$, a choice of dimension $d$ and time lag $\tau$ give rise to a point cloud $\chi = \{\x_i := (x_i,x_{i+\tau}, \cdots,(x_{i+(d-1)\tau}) \} \subset \R^d$.
Then the goal is to analyze this point cloud, which really is a sampling of the full state space, in a way that the dynamics can be understood.
Of course, for practical purposes, not all parameter choices are equally desirable.
While some effort has gone into mathematical justification of ``best'' choices \cite{Casdagli1991}, we are largely left with heuristics that work quite well in practice \cite{Fraser1986,Kennel1992}.

A first method for analyzing this point cloud arose in the form of a recurrence plot \cite{Eckmann1987,Marwan2007}.
Fixing $\e$, this is a binary, symmetric matrix $R = R(\e)$ where $R_{ij}$ is 1 if $\|\x_i-\x_j\| \leq \e$, and 0 otherwise.
Of course, this can be equivalently viewed as the adjacency matrix of a network\footnote{In this paper, we use the words network and graph interchangeably.}, often called the $\e$-recurrence network in this literature \cite{Gao2009,Gao2009a,Marwan2009,Donner2010}.
From this observation, a large literature grew on methods to convert time series into networks; see Donner et al.~\cite{Donner2011} for an extensive survey.

In this paper, we focus on two of these options.
First, given the point cloud $\chi$, we can construct the (undirected) $k$-nearest neighbor graph, commonly called the $k$-NN graph.
This is built by adding a vertex to represent each $\x_i \in \chi$, and for each $\x_i$, adding an edge to the $k$ closest points $\x_j \in \chi$.
This construction, and in particular the investigation of motifs in the resulting graph, has been extensively studied \cite{Shimada2008,Khor2016,Xu2008}.

The second network construction method we work with is the recently developed ordinal partition network \cite{Small2013,McCullough2015}.
It can be viewed as a special case of the class of \textit{transition networks} built from time series data, where vertices represent subsets of the state space, and edges are included based on temporal succession.
This construction arose as a generalization of the concept of permutation entropy \cite{Bandt2002}.
The basic idea of the construction is to replace each $\x = (x_1,\cdots,x_d) \in \chi \subset \R^d$ with a permutation $\pi$ of the set $\{1,\cdots,d\}$ so that $\pi(i) = j$ if $x_j$ is the $i$th entry in the sorted order of the coordinates.
That is, $\pi$ satisfies $x_{\pi(1)} \leq x_{\pi(2)} \leq \cdots \leq x_{\pi(d)}$.
Then we build a graph with vertex set equal to the set of encountered permutations, and an edge included if the ordered point cloud passes from one permutation to the other \footnote{While the ordinal partition network can be defined as a directed, weighted graph, we work with the undirected, unweighted analogue. }.
Utilizing the transition network vantage point, each permutation $\pi$ represents a subspace of $\R^d$ given by the intersection of $\binom{d}{2}$ inequalities, and an edge is included based on passing from one of these subspaces to the other in one time step.
What makes this construction particularly useful is its robustness to noise \cite{Amigo2007}.
See \cite{Amigo2010} for a more extensive introduction.

Many have observed qualitatively that these networks encode the structure of the underlying system \cite{Donner2011}.
In particular, periodic time series tend to create networks with overarching circular structure, while those arising from chaotic systems have more in common with a hairball (see, e.g., \cref{fig:networks_PeriodicVsChaotic}).
However, quantification of this behavior is lacking.
Much of the literature to date has focused on using standard quantification methods from network theory such as local measures like degree-, closeness-, and betweenness-centrality, or the local clustering coefficient.
Global measures are also used, e.g., the global clustering coefficient, transitivity, and average path length.
However, these measures can only do so much to measure the overarching structure of the graph.

It was for this reason that topological data analysis (TDA) \cite{Carlsson2009,Ghrist2014,Ghrist2008,Munch2017,Perea2018} has proven to be quite useful for time series analysis.
TDA is a collection of methods arising from the mathematical field of algebraic topology \cite{Hatcher,Munkres2} which provide concise, quantifiable, comparable, and robust summaries of the shape of data.
The main observation is that we can encode higher dimensional structure than the 1-dimensional information of a network by passing to simplicial complexes.
Like graphs, simplicial complexes are combinatorial objects with vertices and edges, but also allow for higher dimensional analogues like triangles, tetrahedra, etc.
To date, the interaction of time series analysis with TDA has focused on a generalization of the $\e$-recurrence network called a Rips complex and its approximation, the witness complex.
The Rips complex for parameter $\e$ includes a simplex $\sigma = \{ \y_0,\cdots,\y_k\}$ iff $\|\y_i - \y_j\| \leq \e$ for all $i,j$.
That is, it is the largest simplicial complex which has the $\e$-recurrence graph as its 1-skeleton.

Unlike the time series analysis literature, where one works hard to find the perfect $\e$ to construct a single network, TDA likes to work with the Rips complex over all scales in a construction called a filtration.
Then, one can analyze the structure of the overall shape by looking at how long features of interest persist over the course of this filtration.
One particularly useful tool for this analysis is 1-dimensional persistent homology \cite{Edelsbrunner2002,Zomorodian2004}, which encodes how circular structures persist over the course of a filtration in a topological signature called a persistence diagram.
This and its variants have been quite successful in applications, particularly for the analysis of periodicity \cite{Robinson2014,Perea2014,Berwald2014a,Perea2016,Xu2018,Sanderson2017,Robinson2015,Tempelman2019},
including for parameter selection \cite{Garland2016,Maletic2016},
data clustering \cite{Pereira2015},
machining dynamics \cite{Khasawneh2015,Khasawneh2014,Khasawneh2014a,Khasawneh2017,Khasawneh2018b},
gene regulatory systems \cite{Berwald2014b,Perea2015},
financial data \cite{Gidea2017,Gidea2017a,Gidea2018},
wheeze detection \cite{Emrani2014},
sonar classification \cite{Robinson2018},
video analysis \cite{Tralie2016a,Tralie2017,Tralie2018a},
and annotation of song structure \cite{Bendich2018,Tralie2019}.

Unfortunately, while persistence diagrams are powerful tools for summarizing structure, their geometry is not particularly amenable to the direct application of standard statistical or machine learning techniques \cite{Mileyko2011,Turner2014,Munch2015}.
To circumvent the problem, a common trick to deal with persistence diagrams particularly when we are interested in classification tasks is to choose a method for \emph{featurizing} the diagrams; that is, constructing a map from persistence diagrams into Euclidean space $\R^d$ via some method that preserves enough of the structure of persistence diagrams to be reasonably useful.
Many of these exist in the literature \cite{Bubenik2015,Adams2017,Adcock2016,CarlssonVerovsek2016,Kalisnik2018,DiFabio2015,Berry2018,Chevyrev2018,Chen2015a,Chazal2014b,Padellini2017,Perea2019}, however in this work, we focus on the simplest of these realizations, namely point summaries of persistence diagrams which extract a single number for each diagram to be used as its representative.
One summary which we will use in this paper is persistent entropy.
This was defined by Chintakunta et al.~\cite{Chintakunta2015} and later Rucco, Atienza, et al.~\cite{Rucco2017,Atienza2018a} proved that the construction is continuous.
This construction, a modification of Shannon entropy, has found use in several applications \cite{Merelli2015,Rucco2016,Piangerelli2016}.

In this paper we move away from the standard  application of TDA to time series analysis (namely the combination of the Rips complex with 1-D persistence) to implement the following new pipeline and use it to differentiate between chaotic and periodic systems; see \cref{fig:MethodOutline}.
Given a time series, determine a good choice of embedding parameters, and use these to build an embedding of the time series.
Then, obtain a graph either by constructing the $k$ nearest neighbor graph for the points of the embedding, or by  building the ordinal partition network.
Construct a filtration of a simplicial complex using this information, compute its persistence diagram, then return one of several point summaries of the diagram.
We show experimentally on both synthetic and real data that this pipeline, particularly using persistent entropy, is quite good at differentiating between chaotic and periodic time series.
Further, the resulting simplicial complexes used are considerably smaller than those utilized in the Rips complex setting  providing the potential for faster running times.
\section{Background}
\label{sec:background}

\subsection{Graphs}
\label{ssec:Graphs}

%

A graph $G = (V,E)$ is a collection of vertices $V$ with edges $E = \{uv\} \subseteq V \times V$.
In this paper, we assume all graphs are simple (no loops or multiedges) and undirected.
The complete graph on the vertex set $V$ is the graph with all edges included, i.e.~$E = \{uv \mid u \neq v \in V\}$.

We will reference a few special graphs.
The cycle graph on $n$ vertices is the graph $G=(V,E)$ with $V = \{v_1,\cdots,v_n\}$, and $E = \{v_iv_{i+1} \mid 1 \leq i <n \} \cup \{v_nv_1\}$; i.e.~it forms a closed path (cycle) where no repetition occurs except for the starting and ending vertex.
The complete graph on $n$ vertices is the graph $G = (V,E)$ with $V = \{v_1,\cdots,v_n\}$, and $E = \{ v_iv_j \mid i \neq j\}$.
That is, it is the graph with $n$ vertices and all possible edges are included.

We will also work with weighted graphs, $G = (V,E, \omega)$ where $\omega:E \to \R$ gives a weight for each edge in the graph.
In this paper, we assume all weights are non-negative, $\omega: E \to \R_{\geq 0}$.
Given an ordering of the vertices $V = \{v_1,\cdots, v_n\}$, a graph can be stored in an adjacency matrix $\boldA$ where entry $\boldA_{ij} = 1$ if there is an edge $v_iv_j \in E$ and 0 otherwise.
This can be edited to store the weighting information by setting $\boldA_{ij} = \omega(v_iv_j)$ if $v_iv_j \in E$ and 0 otherwise.

A path $\gamma$ in a graph is an ordered collection of non-repeated vertices $\gamma = u_0u_1\cdots u_k$ where $u_iu_{i+1} \in E$ for every $i$.
The length of the path is the number of edges used, namely $\len(\gamma) = k$ in the above notation.
The distance between two vertices $u$ and $v$ is the minimum length of all paths from $u$ to $v$ and is denoted $d(u,v)$. 
Given an ordering of the vertices, this information can be stored in a distance matrix $\boldD$ where $\boldD_{ij} = d(v_i,v_j)$. 
Thus an unweighted graph $G = (V,E)$ gives rise to a weighted complete graph on the vertex set $V$ by setting the weight $\omega(uv) = d(u,v)$.

\subsection{$k$-Nearest Neighbor Graph}
\label{ssec:kNN}
Given a collection of points in $\R^d$, the $k$-nearest neighbor graph, or $k$-NN, is a commonly used method to build a graph.
Fix $k \in \Z_{\geq0}$.
The (undirected) $k$-NN graph has a vertex set in $1$-$1$ correspondence with the point cloud, so we abuse notation and write $v_i$ for both the point $v_i \in \R^d$, and for the vertex $v_i \in V$.
An edge $v_iv_j$ is included if $v_i$ is among the $k$th nearest neighbors of $v_j$.


\subsection{Embedding of time series }
\label{ssec:EmbeddingTimeSeries}

Takens' theorem forms one of the theoretical foundations for the analysis of time series corresponding to nonlinear, deterministic dynamical systems \cite{Takens1981}.
It basically states that in general it is possible to obtain an embedding of the attractor of a deterministic dynamical system from one-dimensional measurements of the system's evolution in time.
An embedding is a smooth map $\Psi: M\to N$ between the manifolds $M$ and $N$ that diffeomorphically maps $M$ to $N$.

Specifically, assume that the state of a system is described for any time $t \in \R$ by a point $\mathbf{x}$ on an $m$-dimensional manifold $M \subseteq \R^d$.
The flow for this system is given by a map $\phi^t(\mathbf{x}): M \times \R \to M$ which describes the evolution of the state $\mathbf{x}$ for any time $t$.
In reality, we typically do not have access to $\mathbf{x}$, but rather have measurements of $\mathbf{x}$ via an observation function $\beta(\mathbf{x}):M  \to \R$.
The observation function has a time evolution $\beta(\phi^t(\x))$, and in practice it is often a one-dimensional, discrete and equi-spaced time series of the form $\{\beta_n\}_{n \in \N}$.

Although the state $\mathbf{x}$ can lie in a higher dimension, the time series $\{\beta_n\}$ is one-dimensional.
Nevertheless, Takens' theorem states that by fixing an embedding dimension $d \geq 2m+1$, where $m$ is the dimension of a compact manifold $M$, and a time lag $\tau > 0$, then the map $\Phi_{\phi,\beta}: M \to \R^d$ given by
\begin{align*}
\Phi_{\phi, \beta} &= (\beta(\mathbf{x}), \beta(\phi(\mathbf{x})), \ldots, \beta(\phi^{d-1}(\mathbf{x}))) \\
                   &= (\beta(\mathbf{x}_t), \beta(\mathbf{x}_{t+\tau}), \beta(\mathbf{x}_{t+2\tau}), \ldots, \beta(\mathbf{x}_{t+(d-1)\tau})),
\end{align*}
is an embedding of $M$, where $\phi^{d-1}$ is the composition of $\phi$ $d-1$ times and $\mathbf{x}_t$ is the value of $\mathbf{x}$ at time $t$.

\begin{figure}
    \includegraphics[width = .48\textwidth]{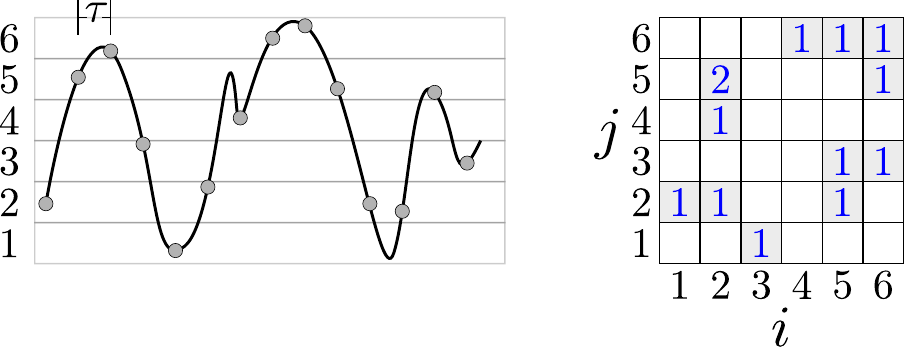}
    \caption{The computation of the mutual information function.
    Each box on the right represents the count of the transitions from the $i$th strip on the left graph to the $j$th strip as the gray points on the time series are traversed from left to right. }
    \label{fig:mutualInfo}
\end{figure}

Theoretically, any time lag $\tau$ can be used if the noise-free data is of infinite precision; however, in practice, the choice of $\tau$ is important in the delay reconstruction.
In this paper we use the first minimum of the mutual information function $I_{\epsilon}(\tau)$ for determining a proper value for $\tau$ \cite{Fraser1986}.
Figure~\ref{fig:mutualInfo} shows pictorially the quantities needed to compute the mutual information function for a fixed $\tau$.
Specifically, the range of the data values is divided into equi-spaced bins with resolution $\epsilon$ and a specific delay value $\tau$ is chosen and fixed.
The joint probability $p_{ij}$ of finding point $\mathbf{x}(t)$ in the $i$th bin and point $\mathbf{x}(t+\tau)$ in the $j$th bin is then computed by counting the number of points laying in the cell indexed by $ij$ in Fig.~\ref{fig:mutualInfo} and dividing this count by the total number of transitions.
In this example, we see that for instance $p_{2\to5}=2/13 \approx 15\%$, while the marginal probability density $p_i$ for $i=5$ is given by $p_{i=5}=3/13\approx23\%$.
Using the probabilities described in Fig.~\ref{fig:mutualInfo}, the mutual information function can be obtained according to
\begin{equation*}
  I_{\epsilon}(\tau) = \sum\limits_{i,j}{p_{ij}(\tau) \ln{p_{ij}(\tau)}} - 2\sum\limits_i{p_i \ln{p_i}},
\end{equation*}
where $\ln(\cdot)$ is the natural logarithm function.
By plotting $I_{\epsilon}(\tau)$ for a range of delays $\tau$, an embedding delay can be chosen by observing the first minimum in $I_{\epsilon}(\tau)$.
This minimum indicates the first value of $\tau$ at which minimum information is shared between $\beta(\mathbf{x}(t))$ and $\beta(\mathbf{x}(t+\tau))$.
We note that the implementation that we describe here is the original implementation described in Kennel et al.~\cite{Kennel1992}; however, an adaptive implementation can be found in Darbellay et al.~\cite{Darbellay1999} while an entropy-based approach can be found in Krasokov et al.~\cite{Kraskov2004}.

The other component in Takens' embedding is the embedding dimension $d$, which must be large enough to unfold the attractor.
If this dimension in not sufficient, then some points can falsely appear to be neighbors at a smaller dimension due to the projection of the attractor onto a lower dimension.
One of the classical time series analysis tools for finding a proper embedding dimension is the method of false nearest neighbors \cite{Kennel1992}.
In this method, it is assumed that an appropriate embedding dimension is given by $d_0$.
Any embedding dimension $d < d_0$ is therefore a projection from $d_0$ into a lower dimensional space.
Consequently, some of the coordinates are lost in this projection and points that are not neighbors in $d_0$ appear to be neighbors in $d$.
The idea is therefore to embed the time series in spaces with increasingly larger dimension while keeping track of false neighbors in successive embeddings, i.e., points that appear to be neighbors due to insufficient embedding dimension.
If the ratio of the false neighbors falls below a certain threshold at some embedding dimension $d$, then we set $d_0\approx d$.

\subsection{Ordinal Partition Graph}
\label{ssec:OrdinalPartitionGraph}
\begin{figure*}
    \centering
    \includegraphics[scale = 4]{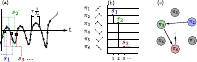}
    \caption{Process for developing ordinal network from times series with permutation dimension $n=3$ and delay $\tau = 3$: (a) permutations from sliding set $s_i = [x_i, x_{i + \tau}, x_{i + 2\tau},... x_{i + (n-1)\tau}]$, (b) Array of ordered permutations $\bf S = [\pi_2, \pi_3, \pi_6... ]$, and (c) directed path in ordinal partition network from $\bf S$.}
    \label{fig:OrdinalPartitionExample}
\end{figure*}

The ordinal partition graph \cite{Small2013,McCullough2015} is another method for constructing a graph from a time series.
Using a sufficiently sampled time series, an ordered list of permutations is collected by first finding a set of $d$-dimensional embedded vectors $v_i$ using a similar algorithm to Takens' embedding.
More specifically, the set of $d$-dimensional vectors  $v_i = [x_i, x_{i + \tau}, x_{i + 2\tau}, \ldots, x_{i + (d-1)\tau}]$ use an embedding delay $\tau$ and motif dimension $d$. An example of this embedding is shown in \cref{fig:OrdinalPartitionExample}~(a) with $\tau = 3$ and $d = 3$.
These parameters were chosen as they simplify the demonstration and visualization of the method.
However, to automate the method, we suggest that both $\tau$ and $d$ are selected by using permutation entropy (PE) as defined by Bandt and Pompe \cite{Bandt2002}. The specific method for selecting $\tau$ and $d$ is explained in Section~\ref{sec:experiments}.
By applying $v_i$ over the entire length of the time series, a sequence of vectors $\bf V$ is generated.

For each vector  $v_i = (x_1,\cdots,x_d)$, the associated permutation $\pi$ is the  permutation of the set $\{1,\cdots,d\}$ that satisfies $x_{\pi(1)} \leq x_{\pi(2)} \leq \cdots \leq x_{\pi(d)}$.
Each vector in $\bf V$ is translated into its associated permutation symbol $\pi_j$ to generate a sequence of permutations $\bf P$, where $j \in \mathbb{Z} \cap [1,n!]$.
An example of this process is shown for the first three vectors in \cref{fig:OrdinalPartitionExample}~(b).
Finally, using the array of permutations $\bf P$, a directional network is formed by transitioning from one permutation, represented by the graph in \cref{fig:OrdinalPartitionExample}~(c), to another in the sequential order.
If we want to build an unweighted version of this graph, we include the edge $\pi\pi'$ if there is a transition from one permutation to the next.
If we want this graph to be weighted, we set $\omega(\pi\pi')$ to be the number of times this transition occurs.

\subsection{Simplicial complexes}
\label{sec:SimplicialComplexes}

A simplicial complex can be thought of as a generalization of the concept of a graph to higher dimensions.
Given a vertex set $V$, a simplex $\sigma \subseteq V$ is simply a collection of vertices.
The dimension of a simplex $\sigma$ is $\dim(\sigma) = |\sigma|-1$.
The simplex $\sigma$ is a face of $\tau$, denoted $\sigma \preceq \tau$ if $\sigma \subseteq \tau$.
A simplicial complex $K$ is a collection of simplices $\sigma \subseteq V$ such that if $\sigma \in K$ and $\tau \preceq \sigma$, then $\tau \in K$.
Equivalently stated, $K$ is a collection of simplices which is closed under the face relation.
The dimension of a simplicial complex is the largest dimension of its simplices, $\dim(K) = \max _{\sigma \in K} \dim(\sigma)$.
The $d$-skeleton of a simplicial complex is all simplices of $K$ with dimension at most $d$,
$K^{(d)} = \{ \sigma \in K \mid \dim(\sigma) \leq d\}$.

Given a graph $G = (V,E)$, we can construct the clique complex
\begin{equation*}
 K(G) = \{ \sigma \subseteq V \mid uv \in E \text{ for all } u\neq v \in \sigma\}.
\end{equation*}
This is sometimes called the flag complex.
The clique complex of the complete graph on $n$ vertices is called the complete simplicial complex on $n$ vertices.

A filtration is a collection of nested simplicial complexes
\begin{equation*}
  K_1 \subseteq K_2 \subseteq \cdots \subseteq K_N.
\end{equation*}
See the bottom row of \cref{fig:ExampleWeightedGraph} for an example of a filtration.
A weighted graph gives rise to a filtration we will make use of extensively.
Given a weighted graph $G = (V,E,\omega)$ and $a \in \R$, we set
\begin{equation*}
  K_a = \{\sigma \in K(G) \mid \omega(uv) \leq a \text{ for all }u \neq v \in \sigma \}.
\end{equation*}
Since $K_a \subseteq K_b$ for $a \leq b$, this can be viewed as a filtration
\begin{equation*}
  K_{a_1} \subseteq K_{a_2} \subseteq \cdots \subseteq K_{a_N}
\end{equation*}
for any collection $a_1 \leq a_2 \leq \cdots \leq a_N$.

In particular, for this paper, we will build a filtration from an unweighted graph $G$ by the following procedure.
First, construct the pairwise distance matrix for the vertices of $G$ using shortest paths.
This can be viewed as a weighting on the complete graph with the same vertex set as $G$.
Thus, it induces a filtration on the complete simplicial complex $K$ where the 1-skeleton of $K_a$ includes edges between any pair of vertices $u$ and $v$ for which $d(u,v) \leq a$.
See \cref{fig:ExampleWeightedGraph} for an example.

\subsection{Homology}
Traditional homology \cite{Hatcher,Munkres2} counts the number of structures of a particular dimension in a given topological space, which in our context will be a simplicial complex.
In this context, the structures measured can be connected components (0-dimensional structure), loops (1-dimensional structure), voids (2-dimensional structure), and higher dimensional analogues as needed.

For the purposes of this paper, we will only ever need 0- and 1-dimensional persistent homology so we provide the background necessary in these contexts.
Further, as a note for the expert, we always assume homology with $\Z_2$ coefficients which removes the need to be careful about orientation.

We start by describing homology.
Assume we are given a simplicial complex $K$.
Say the $d$-dimensional simplices in $K$ are denoted $\sigma_1,\cdots, \sigma_\ell$.
A $d$-dimensional chain is a formal sum of the $d$-dimensional simplices $\alpha = \sum_{i=1}^\ell a_i \sigma_i$.
We assume the coefficients $a_i \in \Z_2 = \{0,1\}$ and addition is performed mod 2.
For two chains $\alpha = \sum_{i=1}^\ell a_i \sigma_i$ and $\beta = \sum_{i=1}^\ell b_i \sigma_i$, $\alpha + \beta = \sum_{i=1}^\ell (a_i + b_i) \sigma_i$.
The collection of all $d$-dimensional chains forms a vector space denoted $C_d(K)$.
The boundary of a given $d$-simplex is
\begin{equation*}
 \partial_d(\sigma) = \sum_{\tau \prec \sigma, \dim(\tau) = d-1} \tau.
\end{equation*}
That is, it is the formal sum of the simplices of exactly one lower dimension.
If $\dim(\sigma) = 0$, that is, if $\sigma$ is a vertex, then we set $\partial_d(\sigma) = 0$.
The boundary operator $\partial_d:C_d(K) \to C_{d-1}(K)$ is given by
\begin{equation*}
\partial_d(\alpha) = \partial_d\left(\sum_{i=1}^\ell a_i \sigma_i \right)
 = \sum a_i \partial_d(\sigma_i).
\end{equation*}

A $d$-chain $\alpha \in C_d(K)$ is a cycle if $\partial_d(\alpha) = 0$; it is a boundary if there is a $d+1$-chain $\beta$ such that $\partial_{d+1}(\beta) = \alpha$.
The group of $d$-dimensional cycles is denoted $Z_d(K)$; the boundaries are denoted $B_d(K)$.

In particular, any $0$-chain is a  $0$-cycle since $\partial_0(\alpha) = 0$ for any $\alpha$.
A $1$-chain is a $1$-cycle iff the 1-simplices (i.e., edges) with a coefficient of 1 form a closed loop.
It is a fundamental exercise in homology to see that $\partial_{d} \partial_{d+1} = 0$ and therefore that $B_d(K) \subseteq Z_d(K)$.
The $d$-dimensional homology group is $H_d(K) = Z_d(K)/B_d(K)$.
An element of $H_d(K)$ is called a homology class and is denoted $[\alpha]$ for $\alpha \in Z_d(K)$ where $[\alpha] = \{ \alpha + \partial(\beta) \mid \beta \in C_{d+1}(K)\}$.
We say that the class is represented by $\alpha$, but note that any element of $[\alpha]$ can be used as a representative so this choice is by no means unique.

In the particular case of 0-dimensional homology, there is a unique class in $H_0(K)$ for each connected component of $K$.
For $1$-dimensional homology, we have one homology class for each ``hole'' in the complex.

\subsection{Persistent homology}
\label{ssec:persistent_homology}

Persistent homology is a tool from topological data analysis which can be used to quantify the shape of data.
The main idea behind persistent homology is to watch how the homology changes over the course of a given filtration.

Fix a dimension $d$.
First, note that if we have an inclusion of one simplicial complex to another $i: K_1 \subseteq K_2$, we have an obvious inclusion map on the $d$-chains $i: C_d(K_1) \to C_d(K_2)$ by simply viewing any chain in the small complex as one in the larger.
Less obviously, this extends to a map on homology $i_*:H_d(K_1) \to H_d(K_2)$ by sending $[\alpha] \in H_d(K_1)$ to the class in $H_d(K_2)$ with the same representative.
That this is well defined is a non-trivial exercise in the definitions \cite{Edelsbrunner2002}.

Given a filtration
\begin{equation*}
  K_1 \subseteq K_2 \subseteq \cdots \subseteq K_N
\end{equation*}
we have a sequence of maps on the homology
\begin{equation*}
  H_d(K_1) \to H_d(K_2) \to \cdots \to H_d(K_N).
\end{equation*}
A class $[\alpha] \in H_d(K_i)$ is said to be born at $i$ if it is not in the image of the map $H_d(K_{i-1}) \to H_d(K_i)$.
The same class dies at $j$ if $[\alpha] \neq 0$ in $H_d(K_{j-1})$ but $[\alpha] = 0$ in $H_d(K_{j})$.

Given all this information, we construct a persistence diagram as follows.
For each class that is born at $i$ and dies at $j$, we add a point in $\R^2$ at $(i,j)$.
For this reason, we often write a persistence diagram by its collection of off-diagonal points, $D = \{(b_1,d_1),\cdots,(b_k,d_k)\}$.
See the top right of \cref{fig:ExampleWeightedGraph} for an example.
Note that the farther a point is from the diagonal, the longer that class persisted in the filtration, which signifies large scale structure.
The \textit{lifetime} or \textit{persistence} of a point $x = (b,d)$ in a persistence diagram $D$ is given by $\pers(x) = |b-d|$.

Note that it is possible to have multiple points in a single diagram of the same form $(b,d)$ if there are multiple classes that are born at $b$ and die at $d$.
In this case, we sometimes employ histograms or annotations to emphasize that a single point seen in a diagram is actually multiple points overlaid.

Further, any filtration has a persistence diagram for each dimension $d$.
In this paper, when we wish to emphasize dimension, we write the diagram as $D_d$; here we will only use the 1-dimensional diagram.

\subsection{A first example}
\label{ssec:FirstExample}
\begin{figure*}
    \centering
    \includegraphics[width = .2\textwidth]{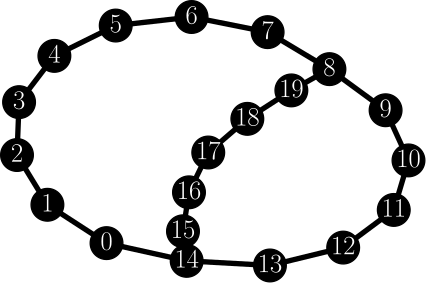}
    \includegraphics[width = .15\textwidth]{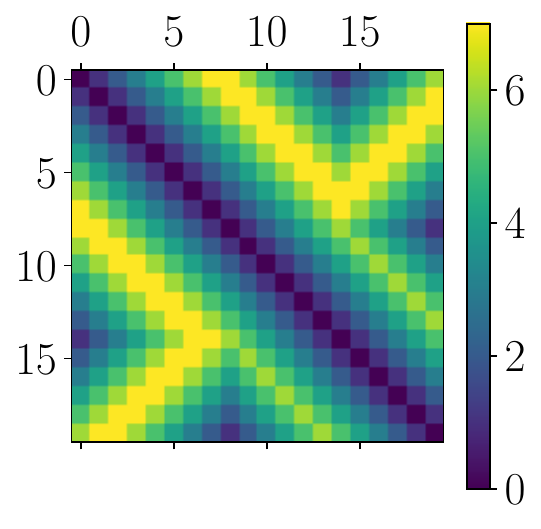}
    \includegraphics[width = .2\textwidth]{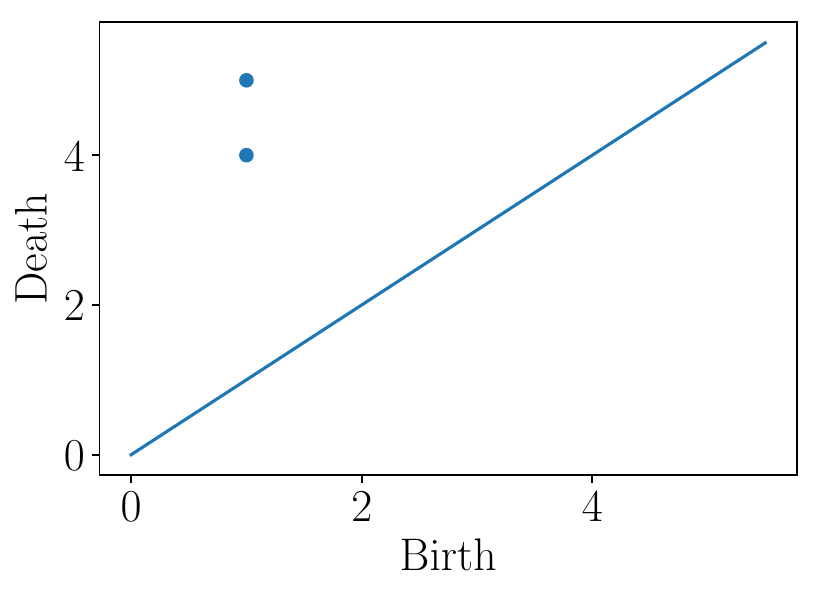}

    \includegraphics[width = .15\textwidth]{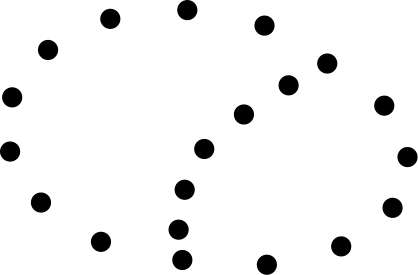}
    \includegraphics[width = .15\textwidth]{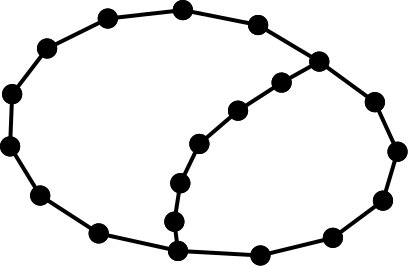}
    \includegraphics[width = .15\textwidth]{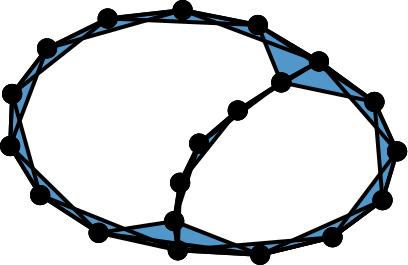}
    \includegraphics[width = .15\textwidth]{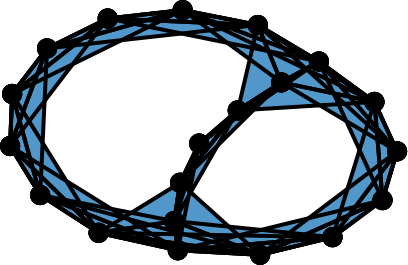}
    \includegraphics[width = .15\textwidth]{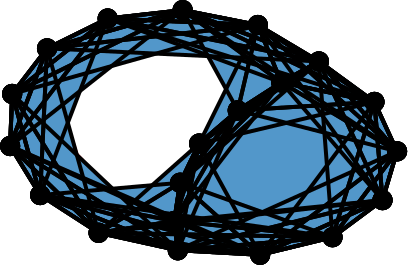}
    \includegraphics[width = .15\textwidth]{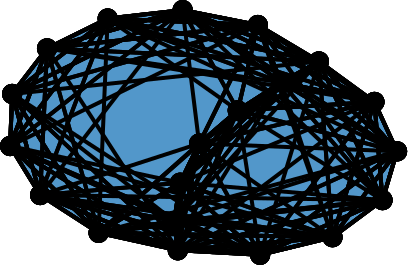}
    \caption{An example of the method used for turning a graph (top left) with pairwise distance information (top middle) into a filtration (bottom row, shown with  thresholded values 0 through 5), and then computing the resulting persistence diagram (top right). }
    \label{fig:ExampleWeightedGraph}
\end{figure*}
Here, we begin to construct the pipeline we will use via an example shown in \cref{fig:ExampleWeightedGraph}.
We start with a graph with $|V|=n=20$ as shown in the top left.
We can construct a distance between each pair of vertices  using the path length.
Then, a filtration on the full simplicial complex $K$ with $n=20$ vertices is constructed using the clique complex method described at the end of \cref{sec:SimplicialComplexes}.
Finally, the 1-dimensional persistence diagram is given at the top right.
The existence of two points in the persistence diagram means that two circular structures existed over the course of the filtration.
The first is the large left loop that can be seen in $K_1$ and persists until it gets filled in at $K_{4}$.
This is represented by the point $(1,4)$ in the diagram.
The other smaller loop is the right loop that appears in $K_1$, but is filled in before $K_{5}$.
This is represented by the point $(1,5)$ in the diagram.

\subsection{Point summaries of persistence diagrams}
\label{ssec:PointSummaryOfPersDgm}

A common issue with persistence diagrams is that they are notoriously difficult to work with as a summary of data.
While they are quantitative in nature, determining differences in structure such as ``has a point far from the diagonal'' is often a qualitative procedure.
Metrics for persistence diagrams exist, namely the bottleneck and $p$-Wasserstein\footnote{This metric is closely related to but not the same as the eponymous metric from probability theory.} distances, however these objects are not particularly easy to work with in a statistical or machine learning context.
Thus, we will pass to working with the simplest of featurizations, namely point summaries of a given diagram, which we also call \textit{scores}.

%
%
%
\paragraph{Maximum persistence}
The first very simple but extremely useful point summary is maximum persistence.
Given a persistence diagram $D$, the maximum persistence is simply
\begin{equation*}
    \maxpers(D) = \max_{x \in D} \pers(x).
\end{equation*}
In the example of \cref{fig:ExampleWeightedGraph} with $D = \{(1,4), (1,5) \}$, we have $\maxpers(D) =  4$.
While this is obviously a very lossy point summary for a persistence diagram, it is quite useful in that, particularly for applications where the existence of a large circle is of interest, it often does what we need.
See, e.g.,  \cite{Khasawneh2015,Tymochko2019}.

\begin{figure}
    \centering
    \includegraphics[scale = 2.4]{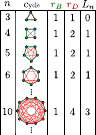}
    \caption{Table of examples showing the lifetime $L_n$ of the single class $(r_B, r_D)$ in the persistence diagram for the pipeline applied to a cycle with $n$ nodes. }
    \label{fig:networks_simplextable}
\end{figure}

\paragraph{Periodicity Score}
Next, we set out to build a point summary which we can use to measure the similarity of our weighted graph to a cycle graph which is independent of the number of nodes.
If $G'$ is an unweighted cycle graph with $n$ vertices, then following the procedure of \cref{ssec:FirstExample} using the shortest path metric, we have that there is exactly one cycle which is born at 1, and fills in at $\lceil \tfrac{n}{3} \rceil$.
See the examples of \cref{fig:networks_simplextable}.
This means the persistence diagram $D'$ has exactly one point $(1,\lceil \tfrac{n}{3} \rceil)$, and so we denote the maximum persistence of this diagram as
\begin{equation*}
    L_n = \maxpers(D') = \left\lceil \frac{n}{3} \right\rceil-1 .
\end{equation*}
Then, assume we are given another unweighted graph $G$ with $|V| = n$ and persistence diagram $D$.
We define the network periodicity score
\begin{equation}
    P(D) = 1-\frac{\maxpers(D)}{L_n}.
    \label{eq:Rn_Equation}
\end{equation}
This score is an extension of the periodicity score in \cite{Perea2015} to unweighted networks, and it has the property that $P(D) \in [0,1]$, with $P(D) = 0$ iff the input graph $G$ is a cycle graph.
In the example of \cref{fig:ExampleWeightedGraph}, we have $L_{20} = 6$, so $P(D) = 1-4/6 = 1/3$.

\paragraph{The ratio of the number of homology classes to the graph order}
The next point summary we define is
\begin{equation}
M(D) = \frac{|D|}{|V|},
\label{eq:persent}
\end{equation}
which is the reciprocal of the ratio between the number of vertices in the network $|V|$, i.e., the order of the graph, and the number of classes in the persistence diagram $|D|$.
In the example of \cref{fig:ExampleWeightedGraph}, this is $M(D) = 2/20 = 0.1$.

We can think of this number as an approximation of the reciprocal of the  number of vertices in each class, however, this is only an approximation because some classes in $1$-D persistence diagram may share vertices in the network.
Note that for a network with $n$ nodes, the 0-dimensional persistence diagram will always have $n-1$ points, and so this metric is not particularly useful.
In this paper, we only use this summary for $1$-dimensional persistence diagrams.

The logic behind this heuristic is that for a periodic signal we would expect to see a small number of $1$-D homology classes in comparison to a chaotic time series.
Therefore, for two networks of similar order but with different dynamic behavior, i.e., one is chaotic and one is periodic, the ratio $M(D)$ for the periodic time series will be smaller than its chaotic counterpart.

\paragraph{Normalized Persistent Entropy}
Persistent entropy is a method for calculating the entropy from the lifetimes of the points in a persistence diagram, inspired by Shannon entropy.
This summary function, first given by Chintakunta et al.~\cite{Chintakunta2015},  is defined as
\begin{equation}
E(D) = - \sum_{x \in D} \frac{\pers(x)}{\mathscr{L}(D)}\log_2\left(\frac{\pers(x)}{\mathscr{L}(D)}\right),
\label{eq:pers_ent}
\end{equation}
where $\mathscr{L}(D) = \sum_{x \in D} \pers(x)$ is the sum of lifetimes of points in the diagram.
In the example of \cref{fig:ExampleWeightedGraph} and \cref{ssec:FirstExample},
$D = \{(1,4), (1,5) \}$, so $\mathscr{L}(D) = 3+4 = 7$.
Thus for this example, $E(D) = 0.985$.

However, we cannot easily compare this value across different diagrams with different numbers of points.
To deal with this issue, we provide the following normalization heuristic.
Specifically, we normalize $E$ as
\begin{equation}
E'(D) = \frac{E(D)}{\log_2\big(\mathscr{L}(D))}.
\label{eq:persent_norm}
\end{equation}
This normalization allows for an accurate measurement of the entropy even when there are few significant lifetimes.
Returning to the example of \cref{fig:ExampleWeightedGraph}, $E'(D) =  0.351$.

\section{Method}
 \label{sec:experiments}

In this section, we discuss the specifics of the method studied for turning a time series into a persistence diagram following \cref{fig:MethodOutline}.

We have two initial choices for how to turn a time series into a network.
In the case of the Takens' embedding, we determine the embedding dimension using false nearest neighbors \cite{Kennel1992}, and determine the lag using the mutual information function \cite{Fraser1986}.
We then construct the $k$-NN graph for these points.
Following Khor and Small~\cite{Khor2016}, we use $k=4$.

The second method for constructing a network is the ordinal partition network.
As mentioned in Section~\ref{ssec:OrdinalPartitionGraph}, this also requires a decision of dimension and lag, which we determine following Reidl et al.~\cite{Riedl2013}.
Specifically, we initially fix $d=3$, and plot the permutation entropy $H(d)=-\sum{p(\pi_i) \log_2{p(\pi_i)}}$, where $p(\pi_i)$ is the probability of a permutation $\pi_i$, for a range of $\tau$ values.
In the resulting $\tau$ versus $H$ curve we choose the value of $\tau$ at the location of the first prominent peak as the lag parameter.
The dimension $d=3$ is used because it was shown in \cite{Riedl2013} that the first peak of $H(d)$ occurs at approximately the same value of $\tau$ independent of the dimension $d$ for $d \in [3, 4, \ldots, 8]$~\cite{Riedl2013}.
The logic behind this approach is that when the time series points are strongly correlated due to the insufficient unfolding of the trajectories, only few regions of the state space are visited resulting in low values for $H$.
As $\tau$ is increased, $H$ increases and it reaches a maximum when the trajectory unfolding leads to the appearance of a large subset of the possible $d!$ motifs.
We only include vertices in the graph for permutations which have been visited, which keeps us from needing to work with the full set of $d!$ which quickly becomes computationally intractable.

Using the identified delay $\tau$ at the first maximum of $H(d=3)$, we then define the permutation entropy per symbol
\begin{equation}{}
    h'(d) = \frac{1}{d-1} H(d),
    \label{eq:norm_PE}
\end{equation}
where we make $d$ a free parameter that we are seeking to determine.
The dimension for the ordinal partition network is obtained by plotting $h'(d)$ for $d \in [3, 4, \ldots, 8]$ and choosing the value of $d$ that maximizes $h'(d)$.

Once the graph is constructed, we compute shortest paths using \texttt{all\textunderscore pairs\textunderscore shortest\textunderscore path\textunderscore length} from the python NetworkX package.
Finally, we compute the persistence diagram using the python wrapper Scikit-TDA \cite{scikittda2019} for the software package Ripser \cite{ripser}.

\subsection{R\"{o}ssler System Example}
\label{ssec:rossler_example}
\begin{figure}
    \centering
    \includegraphics[scale = 0.43]{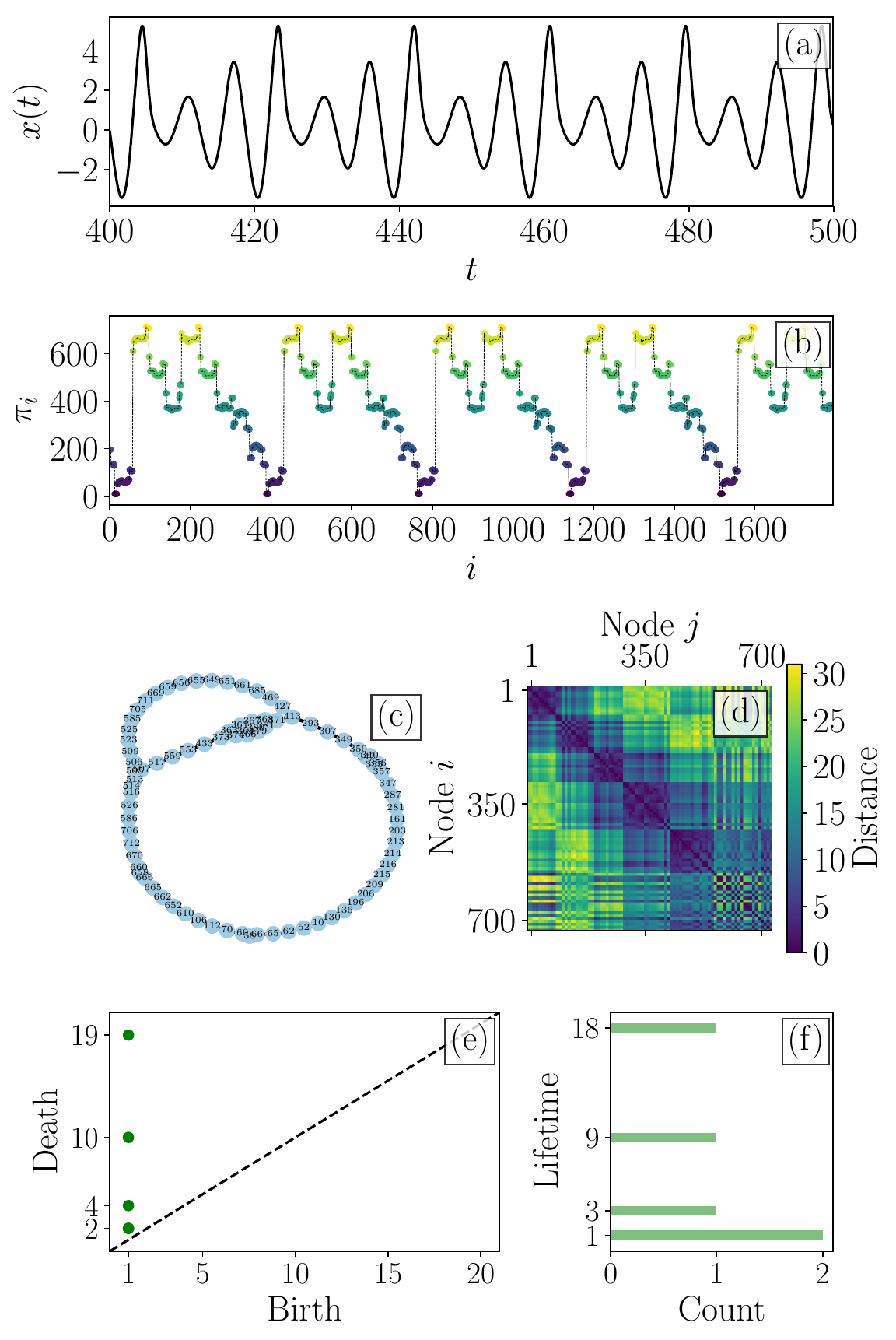}
    \caption{Periodic R\"{o}ssler system example:
    (a) periodic (3-period) time series,
    (b) resulting permutation sequence from embedded time series,
    (c) ordinal partition network drawn with a spring layout,
    (d) pairwise distance matrix using the shortest path metric, and
    (e) the resulting persistence diagram with a histogram (f) showing mulitiplicities of points in the diagram at left.   }
    \label{fig:rossler_ordinalpartition_example}
\end{figure}

We demonstrate the method on the R\"ossler system and the ordinal partition network representation.
The R\"{o}ssler system is defined as
\begin{equation}
\frac{dx}{dt}  = -y - x, \: \frac{dy}{dt}  = x + ay, \: \frac{dz}{dt}  = b + z(x-c).
 \label{eq:rossler}
\end{equation}
Equation~\eqref{eq:rossler} was solved at a rate of 20 Hz for 1000 seconds with parameters $a = 0.41$, $b=2.0$, and $c = 4.0$, which results in a 3-period, periodic response.
Only the last 200 seconds (see Fig.~\ref{fig:rossler_ordinalpartition_example}~(a)) are used to avoid the transients.

We form a permutation sequence from the time series using a time delay $\tau = 40$ and dimension $d = 6$, which were found using MPE as described in Section~\ref{sec:experiments}.
The resulting permutation sequence is shown in Fig.~\ref{fig:rossler_ordinalpartition_example}~(b).
Next, we form the unweighted ordinal partition network shown in Fig.~\ref{fig:rossler_ordinalpartition_example}~(c).
Note that this graph is drawn using the electrical-spring layout function provided by NetworkX since the permutations do not have a natural embedding into Euclidean space.
Using the network, we build the distance matrix in Fig.~\ref{fig:rossler_ordinalpartition_example}~(d).
Finally, by applying persistent homology to the distance matrix, we obtain the persistence diagram in Fig.~\ref{fig:rossler_ordinalpartition_example}~(e).
However, Fig.~\ref{fig:rossler_ordinalpartition_example}~(e) does not show the possibility of point multiplicity in the persistence diagram.
To demonstrate this occurrence we utilize a histogram of the number of classes at each lifetime as shown in Fig~\ref{fig:rossler_ordinalpartition_example}~(f).
This shows there are actually two points in the persistence diagram with lifetime 1.
The point summaries described in Section~\ref{ssec:PointSummaryOfPersDgm} are calculated as $M(D_1) = 0.06$, $P(D_1) = 0.33$, and $E'(D_1) = 0.32$.

\section{Results}
\label{sec:results}
This section compares the persistence-based point summaries and the standard network scores, and illustrates the ability of these scores to detect dynamic state changes.
Specifically, we compare the point summaries $M(D_1)$, $P(D_1)$, and $E'(D_1)$ to some commonly used network quantitative characteristics such as the mean out degree $\langle k \rangle$, the out degree variance $\sigma^2$, and the number of vertices $N$.
These comparisons are shown in Section \ref{ssec:rossler_bifurcation} for a family of trajectories from the R\"{o}ssler system, while Section \ref{ssec:tab_results} tabulates the different scores for a variety of dynamical systems.
In Section~\ref{ssec:effects_Of_Noise} we contrast the noise robustness of our approach to the standard network scores for ordinal partition networks.

\subsection{Dynamic State Change Detection on the R\"{o}ssler System}
\label{ssec:rossler_bifurcation}
Letting the parameter $a$ in \cref{eq:rossler} vary in the range $0.37<a<0.43$ in steps of $\Delta a =0.001$ and setting $\beta=2$ and $\gamma=4$, we obtain $1201$ time series of length $1000$ seconds for the state $x$.
We only retain the last $400$ seconds of the simulation to allow the trajectory to settle on an attractor.
For the construction of the corresponding $k$-NN networks, we sample the time series at $2$ Hz in order to capture a sufficient number of oscillations while avoiding overly large point clouds for computing persistence.
For the Takens' embedding we use the mutual information function approach and the nearest neighbor method, respectively, to choose the parameters $\tau = 4$ and $d = 7$.

For constructing the ordinal partition networks use the higher sampling frequency of $20$ Hz, and we use MPE to select $\tau = 40$ and $d = 6$.
We found that a higher sampling rate for ordinal partition networks and the resulting longer time series is not an issue due to the maximum number of vertices not being dependent on the length of the time series, but rather on the motif dimension $d$ and time series complexity.
Furthermore, a higher sampling rate tends to improve the detection of periodic and chaotic time series for ordinal partition networks.

%
\begin{figure*}
    \centering
    \includegraphics[width = .8\textwidth] {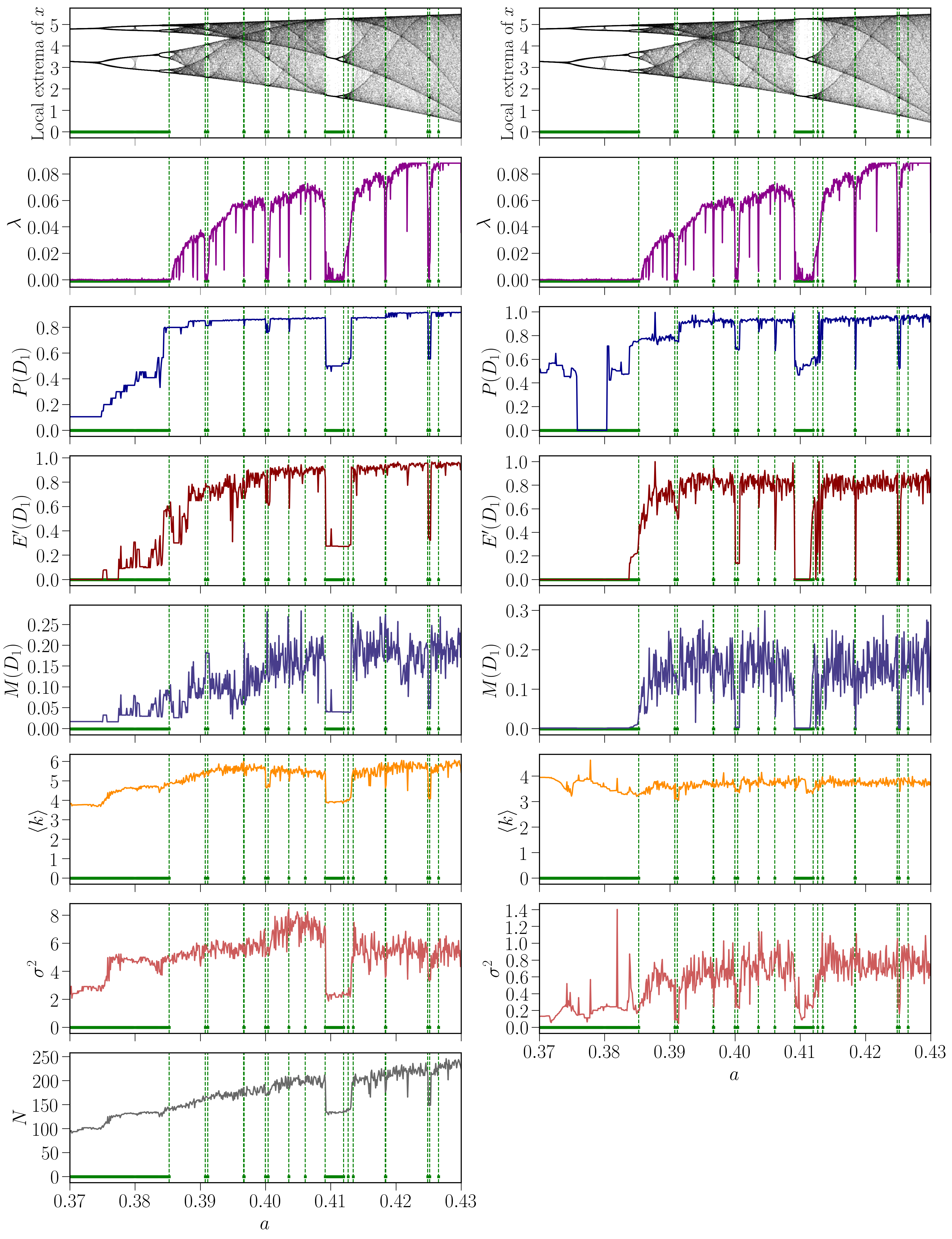}
    \caption{
    R\"{o}ssler system bifurcation for $0.37<a<0.43$ with steps of 0.001 solved using parameters provided in Eq.~\eqref{eq:rossler}.
    Left column plots include point summaries calculated from ordinal partition networks with parameters $\tau = 40$ and $d = 6$;
    Right column plots show the same results for the $k$-NN networks generated from Takens' embedding with parameters $\tau = 4$ and $d = 7$.
    The figure compares point summaries $P(D_1)$, $M(D_1)$, and $E'(D_1)$ with the Lyapunov exponent $\lambda$ \cite{Benettin1980} and some common network parameters including the number of vertices $N$, mean out degree $\langle k \rangle$, and out degree variance $\sigma^2$.
    }
    \label{fig:rossler_Bifurcation_combined}
\end{figure*}

The resulting point summaries were found for both ordinal partition networks (left column plots of Fig.~\ref{fig:rossler_Bifurcation_combined}) and $k$-NN of Takens' embedding networks (right column plots of Fig.~\ref{fig:rossler_Bifurcation_combined}).
The top two graphs in \cref{fig:rossler_Bifurcation_combined} show the bifurcation diagram depicting the local extrema of $x$ and the Lyapunov exponent \cite{Benettin1980}, respectively.
The periodic regions (shown as the regions between vertical,dashed, green lines with a solid green line below) were identified by investigating the bifurcation diagram and the Lyapunov exponent plots.

For the ordinal networks, the left columns plots of Figure~\ref{fig:rossler_Bifurcation_combined} show a significant drop in all six scores for the large periodic window corresponding to approximately $0.409 \leq a \leq 0.412$.
There are also less pronounced drops in these scores for the other shorter periodic windows.
These drops are especially evident for $\langle k \rangle$, $E'(D_1)$, and $P(D_1)$ where the scores significantly decrease in comparison to their surrounding values.
However, some scores such as $\langle k \rangle$ are not normalized, e.g., so that $0 \leq \langle k \rangle \leq 1$.
Given one time series, and not a parameterized set of series, this makes it difficult or even impossible to distinguish between chaotic and periodic regions.
On the other hand, the normalized scores that we introduce in this paper, $E'(D_1)$ and $P(D_1)$, suggest periodic regions when $E'(D_1) < 0.5$ and $P(D_1) < 0.75$.
It should be noted that the difference between chaotic and periodic regions, as shown in Section~\ref{ssec:effects_Of_Noise}, starts degrading as noise levels are increased.

For the $k$-NN Takens' embedding networks, the right column plots of Figure~\ref{fig:rossler_Bifurcation_combined} show a significant drop in $P(D_1)$, $M(D_1)$, and $E'(D_1)$ during periodic windows.
However, for the traditional graph scores $\langle k \rangle$ and $\sigma^2$ this drop does not clearly correspond to the beginning and end of the periodic window.
Further, for the smaller periodic windows interspersed with the chaotic regions we found that $\langle k \rangle$, $\sigma^2$, and $M'(D_1)$ are too noisy to identity the dynamic state changes in these areas.
In contrast, our scores $P(D_1)$ and $E'(D_1)$ retain the ability to distinguish between dynamics regimes, and for $k$-NN networks of Takens' embedding  we suggest tagging the time series as periodic when $E'(D_1) < 0.5$ and $P(D_1) < 0.7$.
%
%
\subsection{Tabulated Results} \label{ssec:tab_results}
This section uses a variety of dynamical systems to validate the observations we made for the R\"{o}ssler system in \cref{ssec:rossler_bifurcation} related to the point summaries $E'(D_1)$, $M(D_1)$, and $P(D_1)$ that we introduced in \cref{ssec:PointSummaryOfPersDgm}.
The results for each system when using ordinal partition networks and the $k$-NN network from Takens' embedding are provided side by side in Table~\ref{tab:point_summary_comparisons}.
The model and time series information for all of these systems are provided in \cref{sec:appx:DynamicalSystems}.
The table can be categorized into three types of dynamical systems: (1) systems of differential equations (Chua circuit, Lorenz, R\"{o}ssler, coupled Lorenz-R\"{o}ssler, bi-directional R\"{o}ssler, and Mackey-Glass equations), (2) discrete-time dynamical systems (Logistic map, and H\'{e}non map), and (3) ECG and EEG signals.
The paragraphs below discuss the results for each one of these systems.
\begin{center}
\begin{table*}
\centering
\begin{tabular}{|c|c|c|c|c|c|c|c|c|c|c|c|c|c|}
\hline
\multirow{3}{*}{\textbf{\begin{tabular}[c]{@{}c@{}}System/\\   Data\end{tabular}}} & \multirow{3}{*}{\textbf{Ref.}} & \multicolumn{6}{c|}{\textbf{\begin{tabular}[c]{@{}c@{}}$k$-NN Graph from \\ Takens' Embedding\end{tabular}}} & \multicolumn{6}{c|}{\textbf{Ordinal Partition Graph}} \\ \cline{3-14}
 &  & \multicolumn{2}{c|}{\textbf{$E'(D_1)$}} & \multicolumn{2}{c|}{\textbf{$M(D_1)$}} & \multicolumn{2}{c|}{\textbf{$P(D_1)$}} & \multicolumn{2}{c|}{\textbf{$E'(D_1)$}} & \multicolumn{2}{c|}{\textbf{$M(D_1)$}} & \multicolumn{2}{c|}{\textbf{$P(D_1)$}} \\ \cline{3-14}
 &  & \textbf{Per.} & \textbf{Ch.} & \textbf{Per.} & \textbf{Ch.} & \textbf{Per.} & \textbf{Ch.} & \textbf{Per.} & \textbf{Ch.} & \textbf{Per.} & \textbf{Ch.} & \textbf{Per.} & \textbf{Ch.} \\ \hline
Chua Circuit & \ref{app:chua} & 0.00 & 0.80 & 0.001 & 0.19 & 0.54 & 0.89 & 0.21 & 0.72 & 0.051 & 0.19 & 0.42 & 0.88 \\ \hline
Lorenz & \ref{app:lorenz} & 0.04 & 0.84 & 0.005 & 0.16 & 0.64 & 0.93 & 0.18 & 0.95 & 0.026 & 0.36 & 0.28 & 0.96 \\ \hline
Rossler & Eq.~\eqref{eq:rossler} & 0.00 & 0.85 & 0.001 & 0.18 & 0.50 & 0.94 & 0.00 & 0.89 & 0.036 & 0.28 & 0.33 & 0.85 \\ \hline
\begin{tabular}[c]{@{}c@{}}Coupled \\ Lorenz-Rossler\end{tabular} & \ref{app:lorenz_rossler} & 0.00 & 0.82 & 0.003 & 0.16 & 0.46 & 0.94 & 0.00 & 0.87 & 0.033 & 0.35 & 0.56 & 0.92 \\ \hline
\begin{tabular}[c]{@{}c@{}}Bi-directional \\ Rossler\end{tabular} & \ref{app:bi_rossler} & 0.00 & 0.76 & 0.004 & 0.13 & 0.55 & 0.87 & 0.25 & 0.91 & 0.064 & 0.29 & 0.40 & 0.92 \\ \hline
Mackey-Glass & \ref{app:mackey_glass} & 0.00 & 0.67 & 0.001 & 0.07 & 0.56 & 0.93 & 0.30 & 0.96 & 0.077 & 0.37 & 0.25 & 0.93 \\ \hline
Logistic Map & \ref{app:logistic} & \multicolumn{6}{c|}{\multirow{2}{*}{NA}} & 0.00 & 0.93 & 0.125 & 0.70 & 0.00 & 0.91 \\ \cline{1-2} \cline{9-14}
Henon Map & \ref{app:henon} & \multicolumn{6}{c|}{} & 0.00 & 0.88 & 0.111 & 0.48 & 0.00 & 0.96 \\ \hline
ECG & \ref{app:ecg} & 0.95 & 0.86 & 0.282 & 0.14 & 0.97 & 0.97 & 0.82 & 0.89 & 0.268 & 0.45 & 0.92 & 0.97 \\ \hline
EEG & \ref{app:eeg} & 0.96 & 0.94 & 0.627 & 0.33 & 0.99 & 0.98 & 0.89 & 0.84 & 0.513 & 0.31 & 0.97 & 0.93 \\ \hline
\end{tabular}
\caption{A comparison between persistence diagram point summaries $M(D_1)$, $P(D_1)$, and $E'(D_1)$ for detecting differences in the networks generated from for periodic (Per.) and chaotic (Ch.) time series using both $k$-NN graphs and ordinal partition graphs.}
\label{tab:point_summary_comparisons}
\end{table*}
\end{center}

\paragraph{Systems of differential Equations:}
As shown in Table~\ref{tab:point_summary_comparisons}, our point summaries from both networks yield distinguishable differences between periodic and chaotic time series.
The $k$-NN graph results in Table~\ref{tab:point_summary_comparisons} show that periodic time series have $E'(D_1) < 0.5$, $M(D_1) < 0.15$, and $P(D_1) < 0.7$.
Similarly, the ordinal partition graph scores in Table~\ref{tab:point_summary_comparisons} show that periodic time series have $E'(D_1) < 0.5$, $M(D_1) < 0.07$, and $P(D_1) < 0.75$.

\paragraph{Discrete dynamical systems:}
The results for the discrete dynamical equations in Table~\ref{tab:point_summary_comparisons} show distinguishable differences between periodic maps in comparison to chaotic maps when using ordinal partition networks.
Takens' embedding was not applied to the discrete dynamical systems, and only the ordinal partition network results are reported here because working with these networks is more natural for maps.

\paragraph{EEG and ECG Results:}
The point summary results from real world data sets (ECG and EEG) shown in Table~\ref{tab:point_summary_comparisons} have inherent noise, which causes the differences between the compared states to be less significant as shown in Fig.~\ref{fig:SNRsweep_scores_ordinalNet}.
The $k$-NN graph results in Table~\ref{tab:point_summary_comparisons} do not show a significant difference between the two groups for either ECG and EEG data.
This is most likely due to the sensitivity of Takens' embedding to noise and perturbations.
However, we did find a difference between epileptic and healthy patients through the networks formed by ordinal partitions for ECG~\cite{Moody1992} and EEG~\cite{Andrzejak2001} data.
\cref{ssec:effects_Of_Noise} discusses the effect of additive noise on the point summaries in more detail.
As a note, there have been other methods for characterizing EEG data using TDA and persistent entropy~\cite{Piangerelli2016}, but our method is different from prior works because we apply persistent homology to the generated networks.

\subsection{Effects of Additive Noise}
\label{ssec:effects_Of_Noise}
In this section we investigate the noise robustness of the point summaries in comparison to some common network parameters---mean out degree $\langle k \rangle$, out degree variance $\sigma^2$, and the number of vertices $N$.
The ordinal partition networks are based on time series from the R\"{o}ssler system defined in Eq.~\eqref{eq:rossler} with $b=2.0$, $c = 4.0$, and either $a = 0.41$ or $a = 0.43$ for a periodic or chaotic response, respectively.

To make comparisons on the noise robustness we add Gaussian noise to the signal and calculate the point summaries and network parameters at various Signal-to-Noise Ratios (SNR) for both periodic and chaotic R\"{o}ssler systems.
The chosen SNR values were all the integers from $1$ to $50$, and at each SNR value we obtain $25$ realizations of noisy signals.

\begin{figure}
    \centering
    \includegraphics[scale = 0.41]{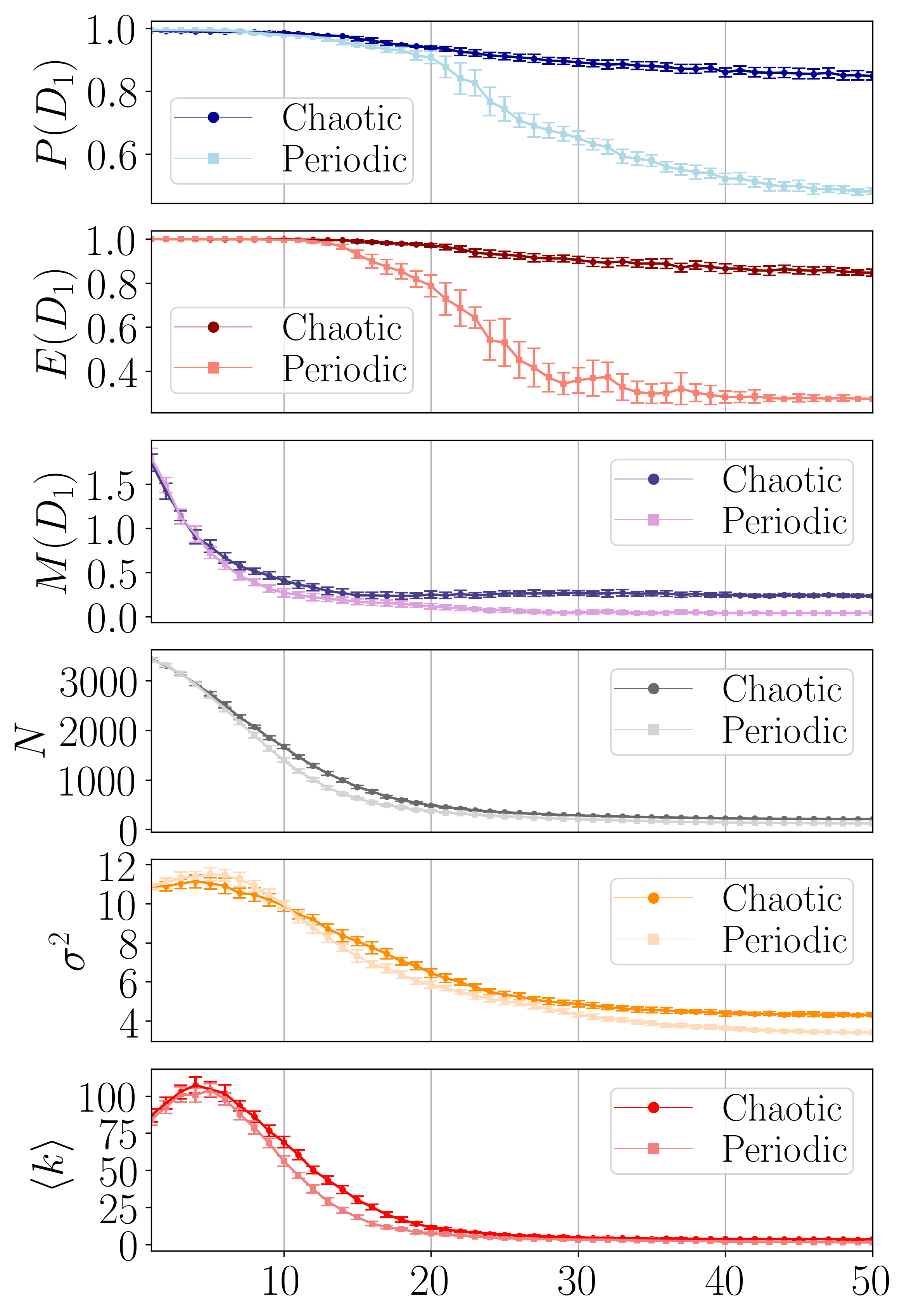}
    \caption{Average point summaries and network parameters for varying SNRs from Gaussian noise added to time series generated from periodic and chaotic R\"{o}ssler systems. For each SNR, 25 separate samples are taken to provide mean values and standard deviations, which are shown as the error bars.}
    \label{fig:SNRsweep_scores_ordinalNet}
\end{figure}

To determine the $68\%$ confidence interval at each SNR, we repeat the calculation of the point summaries and network parameters for all noise realizations at each SNR level, and we set our confidence interval to $\overline{x}(SNR) \pm s(SNR)$ where $\overline{x}(SNR)$ and $s(SNR)$ are the sample average and sample standard deviation, respectively, at a specific SNR value.
Figure~\ref{fig:SNRsweep_scores_ordinalNet} shows the mean values and confidence intervals for each SNR.
To assess the ability of point summaries to assign a distinguishing score to a periodic versus a chaotic system in the presence of noise, we check for an overlap in the confidence intervals for the periodic and chaotic results at each SNR.
If for a particular point summary there is an overlap between the scores for periodic and the chaotic time series, then that point summary is not effective in distinguishing the dynamics at that specific SNR.
Table~\ref{tab:noise_robustness_table} summarizes the noise robustness by providing the lowest SNR at which each point summary and network parameter no longer has an overlap between the periodic and chaotic confidence intervals.
This result shows a lower capable SNR for the persistence based point summaries than the mean out degree $\langle k \rangle$ and variance $\sigma^2$.
Another trend that should be noted is the reduction in difference between periodic and chaotic time series for high levels of noise. This should be taken into account when applying the point summaries to real world data with intrinsic noise.

\begin{table}
\begin{tabular}{|c|c|}
\hline
\begin{tabular}[c]{@{}c@{}}Point Summary/\\ Network Parameter\end{tabular} & Lowest Distinguishing SNR \\ \hline
$E'(D_1)$ & 14 \\ \hline
$M(D_1)$ & 19 \\ \hline
$P(D_1)$ & 20 \\ \hline
$\langle k \rangle$ & 29 \\ \hline
$\sigma^2$ & 29 \\ \hline
$N$ & 8 \\ \hline
\end{tabular}
\caption{Noise robustness comparison for persistence diagram point summaries and network parameters using ordinal partition network. }
\label{tab:noise_robustness_table}
\end{table}
%
%
\section{Conclusions}
\label{sec:conclusions}
In this paper we develop a  new framework for time series analysis using TDA.
We investigate two methods for embedding a time series into an unweighted graph: (1) utilizing standard Takens' theorem techniques, then building a $k$-NN graph; and (2) using ordinal partition networks to turn (visited) parts of the state-space into symbols, and obtaining a graph by tracking sequential transitions between these subspaces.
We then describe how to obtain the $1$-D persistence diagram corresponding to the graph by defining a filtration on the full simplicial complex using the pairwise distances between the graph vertices.
The obtained persistence diagram then allows the application of tools from TDA to gain insights into the system's underlying dynamics.
Specifically, a graph embedding of a periodic time series is long connected network loops, while a chaotic time series has many short loops.
These characteristics allow persistent homology to accurately distinguish periodic and chaotic time series by measuring the shape of the networks.

In addition to describing this novel approach for time series analysis, another contribution of this work is the introduction of new point summaries for extracting information about the dynamic state (periodic or chaotic) from time series measurements.
Specifically, we extend the periodicity score $P(D_1)$, which was defined on $\R^n$ in Ref.~\cite{Perea2015}, to abstract graph spaces.
We also define a heuristic $M(D_1)$ which represents an approximation of the ratio of the number of homology classes to the graph order.
The last point summary we define is a normalized version of the persistence entropy $E'(D_1)$ \cite{Chintakunta2015}.

We found that these point summaries outperform standard graph scores, see Fig.~\ref{fig:rossler_Bifurcation_combined}.
Specifically, our point summaries are more capable of distinguishing shifts in the dynamic behavior than their traditional graph scores counterpart.
Further our point summaries, especially the two normalized scores $P(D_1)$ and $E'(D_1)$, enable making inferences about the dynamic behavior from isolated time series, as opposed to tracking changes in the scores of parameterized time series some of which belong to a known dynamic regime.
For example, applying our point summaries to ordinal partition networks from a variety of dynamical systems in Table~\ref{tab:point_summary_comparisons}, we observed that that periodic time series typically have $E'(D_1) < 0.5$, $M(D_1) < 0.15$, and $P(D_1) < 0.7$.
Similarly, using the networks obtained from $k$-NN embedding shows that periodic time series have $E'(D_1) < 0.5$, $M(D_1) < 0.07$, and $P(D_1) < 0.75$, see Table~\ref{tab:point_summary_comparisons}.
However, for both discrete dynamical systems as well as ECG and EEG data, only the persistent homology of the ordinal partition network was able to distinguish between the two data sets. 
Additionally, we showed in \cref{fig:SNRsweep_scores_ordinalNet} that the point summaries of the ordinal partition networks are noise robust down to an SNR of approximately $15$ with additive Gaussian noise.
In future work, to develop more precise ranges for periodic point summary scores and to improve the dynamic state detection, it would be beneficial to investigate the statistical significance of these point summaries as well as the correlation between the point summaries for the $k$-NN and ordinal partition networks.
\section*{Acknowledgments}
\label{sec:acknowledgments}
The authors thank the anonymous reviewers for helpful feedback. 
This material is based upon work supported by the National Science Foundation under grant
nos. CMMI-1759823 and DMS-1759824 with PI FAK, and CMMI-1800466 and DMS-1800446 with PI EM.
\bibliography{ordinal_plus_TDA}
\newpage
\appendix
\section{Dynamical System Examples}
\label{sec:appx:DynamicalSystems}
\subsection{Chua Circuit}
\label{app:chua}
The Chua Circuit used is defined as
	\begin{equation}
	\begin{split}
	\frac{dx}{dt}  &  = \alpha (y-x-f(x)), \\
	\frac{dy}{dt}  &  = \gamma (x-y+z), \\
	\frac{dz}{dt}  &  = - \beta y,
	\end{split}
 	\label{eq:chua}
	\end{equation}
where $f(x)$ is defined as $f(x) = m_1 x + \frac{1}{2}(m_0 +m_1)(|x+1|-|x-1|)$, with $m_0 = -8/7$ and $m_1=-5/7$.
Additionally, the Chua circuit had a sampling rate of 50 Hz with parameters $\alpha = 15.6$, $\gamma = 1.0$, and $\beta = 33.80$ for a periodic response or $\beta = 33.55$ for a chaotic response. This system was solved for 500 seconds and the last 100 seconds were used. The generated time series were downsampled to 7 Hz for $k$-NN networks.
\subsection{Lorenz System}
\label{app:lorenz}
The Lorenz system used is defined as
\begin{equation}
\frac{dx}{dt}   = \sigma (y-x), \: \frac{dy}{dt}   = x (\rho -z) - y, \: \frac{dz}{dt}   = xy - \beta z.
 \label{eq:lorenz}
\end{equation}
The Lorenz system had a sampling rate of 100 Hz with parameters $\sigma = 10.0$, $\beta = 8.0 / 3.0$, and $\rho = 180.1$ for a periodic response or $\rho = 181.0$ for a chaotic response. This system was solved for 100 seconds and the last 24 seconds were used. The generated time series were downsampled to 35 Hz for $k$-NN networks.
\subsection{Coupled Lorenz-R\"{o}ssler System}
\label{app:lorenz_rossler}
The coupled Lorenz-R\"{o}ssler system used is defined as
    \begin{equation}
    \begin{split}
	\frac{dx_1}{dt}  &  = -y_1 - z_1 + k_1 (x_2 - x_1), \\
	\frac{dy_1}{dt}   & = x_1 + a_2 y_1 + k_2 (y_2 - y_1), \\
	\frac{dz_1}{dt}  &  = b_2 + z_1 (x_1 - c_2) + k_3 (z_2 - z_1), \\
	\frac{dx_2}{dt}  &  = \sigma (y_2-x_2), \\
	\frac{dy_2}{dt} &  = \lambda x_2 - y_2 - x_2 z_2, \\
	\frac{dz_2}{dt}   & = x_2 y_2 - b_1 z_2,
	\end{split}
 	\label{eq:lorenz_lorenz}
	\end{equation}
with $\sigma = 10$, $b_1 = 8/3$, $b_2 = 0.2$, $c_2 = 10$, $k_1 = 1$, $k_2 = 10$, $k_3 = 0$, $\lambda = 28$, and $a_2 = 0.25$ for a periodic response or $a_2 = 0.51$ for a chaotic response. This was solved for 400 seconds with a sampling rate of 50 Hz. Only the last 200 seconds of the x-solution were used in the analysis. The generated time series were downsampled to 2 Hz for $k$-NN networks.

\subsection{Bi-Directional Coupled R\"{o}ssler System}
\label{app:bi_rossler}
The Bi-directional R\"{o}ssler system is defined as
    \begin{equation}
    \begin{split}
	\frac{dx_1}{dt} & = -w_1y_1 - z_1 + k(x_2-x_1), \: \\
	\frac{dy_1}{dt} & = w_1x_1 + 0.165y_1, \:	\\
	\frac{dz_1}{dt} & = 0.2 + z_1(x_1-10), \\
	\frac{dx_2}{dt} & = -w_2y_2 - z2 + k(x_1-x_2), \\
	\frac{dy_2}{dt} & = w_2x_2 + 0.165y_2, \:	 \\
	\frac{dz_2}{dt} & = 0.2 + z_2(x_2-10), \\
	\end{split}
 	\label{eq:rossler_rossler}
	\end{equation}
with $w_1 = 0.99$, $w_2 = 0.95$, and $k = 0.0544$ for a periodic response or $k = 0.0558$ for a chaotic response. This was solved over 4000 seconds with a sampling rate of 10 Hz. Only the last 400 seconds of the x-solution were used in the analysis. The generated time series were downsampled to 1 Hz for $k$-NN networks. The generated time series were downsampled to 2 Hz for $k$-NN networks.
\subsection{Mackey-Glass Delayed Differential Equation}
\label{app:mackey_glass}
The Mackey-Glass Delayed Differential Equation is defined as
    \begin{equation}
	x(t) = -\gamma x(t) + \beta \frac{x(t-\tau)}{1+{x(t-\tau)}^n}
 	\label{eq:Mackey_Glass}
	\end{equation}
with $\tau = 2$, $\beta = 2$, $\gamma = 1$, and $n = 7.00$ for a periodic response or $n = 9.65$ for a chaotic response. This was solved for 400 seconds with a sampling rate of 100 Hz. Only the last 300 seconds of the x-solution were used in the analysis. The generated time series were downsampled to 2.5 Hz for $k$-NN networks and 25 Hz for ordinal partition networks.
\subsection{EEG Data}
\label{app:eeg}
The EEG signal was taken from andrzejak et al.~\cite{Andrzejak2001}. More specifically, the first 2000 data points from the EEG data of a healthy patient from set A, file Z-093 was used as the periodic series, and the first 2000 data points from the EEG data of a patient during active seizures from set E, file S-056 was used as the chaotic series. The generated time series were downsampled to 80 Hz for $k$-NN networks.
\subsection{ECG Data}
\label{app:ecg}
The Electrocardoagram (ECG) data was taken from SciPy's misc.electrocardiogram data set. This ECG data was originally provided by the MIT-BIH Arrhythmia Database~\cite{Moody1992}. We used data points 3000 to 4500 during normal sinus rhythm as the periodic time series and data points 8500 to 10000 during ventricular contractions as the chaotic time series.
\subsection{Logistic Map}
\label{app:logistic}
The logistic map was generated as
	\begin{equation}
   	x_{n+1} = r x_n(1-x_n),
   	\label{eq:log_map}
	\end{equation}
with $x_0 = 0.5$ and $r = 3.50$ for periodic results or $r = 3.95$ for chaotic results. Equation~\ref{eq:log_map} solved for the first 500 data points.
\subsection{H\'{e}non Map}
\label{app:henon}
The H\'{e}non map was solved as
	\begin{equation}
	\begin{split}
	x_{n+1} & = 1 - a x_n^2 + y_n, \\
	y_{n+1} & = b x_n,
	\end{split}
 	\label{eq:henon_map}
	\end{equation}
where $b = 0.3$, $x_0 = 0.1$, $y_0 = 0.3$, and $a = 1.05$ for a periodic response and $a = 1.4$ for a chaotic response. This system was solved for the first 500 data points of the x-solution.
\end{document}